\documentclass[conference]{IEEEtran}
\usepackage{graphics}
\usepackage{graphicx}
\usepackage{cite}
\usepackage{xcolor}
\usepackage[]{algorithm2e}
\usepackage{listings}
\usepackage{float}
\usepackage{amssymb,amsfonts,pifont}
\usepackage[bookmarks=true,%
            bookmarksnumbered=true,%
            colorlinks=true,%
            linkcolor=linkcol,%
            citecolor=citecol,%
            urlcolor=urlcol,%
            hypertexnames=true,%
            pdfpagelabels]%
      {hyperref}
   \hypersetup{ pdfauthor = {Jeremiah Onaolapo},
                pdftitle = {},
                pdfkeywords = {},
                pdfcreator = {},
                pdfproducer = {}
              }

   \definecolor{linkcol}{rgb}{0,0,0.5}
   \definecolor{citecol}{rgb}{0,0.5,0.3}
   \definecolor{urlcol}{rgb}{0.3,0,0}
\usepackage{url}
\usepackage{todonotes}
\usepackage{amsfonts}
\usepackage{amstext}
\usepackage[font=footnotesize]{subfig}
\usepackage{fixltx2e}
\usepackage{multirow}

\newcommand{\hyp}[1]{$H_{#1}$} 
\newcommand{\hypl}[1]{Hypothesis #1} 
\newcommand{\desc}[1]{\noindent \textbf{#1.}} 

\title{All Your Cards Are Belong To Us: \\Understanding Online Carding Forums}
\author{Andreas Haslebacher, Jeremiah Onaolapo, and Gianluca Stringhini\\
  University College London\\
\texttt{andreas.haslebacher.14@ucl.ac.uk}\\ \texttt{\{j.onaolapo,g.stringhini\}@cs.ucl.ac.uk}}

\begin{document}

\maketitle
\thispagestyle{empty}

\begin{abstract}
Underground online forums are platforms that enable trades of illicit services and stolen goods. Carding forums, in particular, are known for being focused on trading financial information. However, little evidence exists about the sellers that are present on active carding forums, the precise types of products they advertise, and the prices buyers pay. Existing literature mainly focuses on the organisation and structure of the forums. Furthermore, studies on carding forums are usually based on literature review, expert interviews, or data from forums that have already been shut down. This paper provides first-of-its-kind empirical evidence on active forums where stolen financial data is traded. We monitored five out of 25 discovered forums, collected posts from the forums over a three-month period, and analysed them quantitatively and qualitatively. We focused our analyses on products, prices, seller prolificacy, seller specialisation, and seller reputation, and present a detailed discussion on our findings.


\end{abstract}

\section{Introduction}

\pounds479 million of fraud losses on UK issued credit and debit cards were recorded in 2014~\cite{ffauk2015}. Almost 70\% of these losses originate from ``remote purchase fraud.'' This category of fraud denotes that card details obtained through illicit methods such as phishing, skimming or hacking are used for fraudulent online transactions. Since consumers have mostly shifted from cash to transactions via payment cards and have become accustomed to online payments, opportunities for theft of payment card details have soared and they have attracted the attention of cybercriminals. 

Theft of card information is usually the first step in the chain of credit card fraud. Further stages are resale, validation and monetization of the stolen data. These deals and activities take place in a massive underground economy, usually aided by underground online forums. These forums are popular platforms where card details are traded, thus generating huge revenues for cybercriminals. On these forums, fraudsters typically open a thread and write an advertisement for their products as a first posting. Potential buyers either reply within that thread asking to contact them or they contact the seller themselves using private message services or instant messaging services like ICQ. 

The sales volumes thus generated appear to be substantial. It is estimated, for example, that the closure of several credit card related forums in 2012 prevented international fraud to the tune of \pounds500 million~\cite{gold2013identity}. It is therefore important to understand the characteristics of these online forums and the activity of cybercriminals using them. 

The body of research into underground forums is growing but still limited. In particular, there are only a few studies available about credit card related forums. These studies mainly focus on the organisation and the structure of the forums but less on the content itself, that is, the products traded and the activity of the traders on these forums~\cite{allodi2015then,afroz2013honor,yip2013forums}. 
In addition, existing studies are usually based on either expert interviews or examinations of forums that have been shut down by law enforcement agents. Active forums are rarely analysed. The examination of closed forums may be problematic since they may differ from those still existing, especially when this difference is the reason why they are closed. Moreover, cybercrime evolves rapidly and tackling this type of crime requires an understanding of the current situation and activities.

In this paper, we collected data directly from the discussions on the underground forums that we studied, with emphasis on product offers and advertisements posted by potential sellers on the forums. Since sellers come earlier in the fraud chain than the other actors, it might be more efficient to tackle credit card fraud by stopping sellers than buyers, for example. As a result, we excluded buyers and money mules from this study (we will study them in future work).

\desc{Research question} In order to overcome the literature gaps and methodological concerns we highlighted earlier, this paper aims to shed light on the current situation of underground online forums by analysing real data collected from active forums. We set out to answer the following research question: What does the current situation of underground credit card forums look like? That is, what are the typical features of the forums, what products are sold and how can the activity of the traders be characterised? 

\desc{Hypotheses} After studying existing research literature, we defined the following hypotheses to guide our analyses:

\begin{itemize}

	\item \hypl{1} (\hyp{1}). Prices of credit card numbers combined with additional information about the cardholder are higher than prices of credit card numbers alone.

	\item \hypl{2} (\hyp{2}). On active carding forums, a small number of traders are responsible for a large proportion of traffic.

	\item \hypl{3} (\hyp{3}). Specialisation is discernible on carding forums, that is, most of the traders sell only one product type.
  
	\item \hypl{4} (\hyp{4}). Specialised traders sell their products at lower prices than unspecialised traders.

	\item \hypl{5} (\hyp{5}). The carding forums under analysis have working reputation systems that are sophisticated as those of legal marketplaces.

	\item \hypl{6} (\hyp{6}). The vast majority of actors are not operating on more than one forum.

\end{itemize}

Our analyses confirmed~\hyp{1}, \hyp{2}~and \hyp{6}. \hyp{4} was partially rejected, while \hyp{3}~and \hyp{5}~were completely rejected. Details of the analyses are in Section~\ref{sec:analysis}.

\desc{Contributions} In this paper, we established an outline of active forums and their defining features. To this end, we analyzed five carefully selected active forums in detail regarding their traders and products. This investigation thus provides insights into current underground online forums, unlike previous work that studied forums that were shut down. This paper provides some insights into product proportions on carding forums, since existing literature does not provide any sound information on that. Overall, we present a comprehensive overview of carding forums. We identified various product types with different prices. Our findings suggest that a small number of traders are responsible for the majority of the traffic observed on the underground forums. A distinct pattern of seller specialisation is not yet discernible from our findings.

\section{Background and Related Work}

In this section, we explore products sold on underground forums, their prices, seller prolificacy, seller specialization, and seller reputation, as presented in previous work. In addition, we develop some hypotheses from the general findings presented in existing literature. These hypotheses form the basis of our analysis, presented later in this paper. 

\subsection{Products and prices}
Products and services traded on carding forums can generally be classified as credit card information, bank account information, credentials or online payment services~\cite{ablon2014markets}. 
In 2011, Motoyama et al.~\cite{motoyama2011analysis} analysed the records of six closed forums and identified online payments, game-related accounts, credit cards and financial accounts as being the items most traded.
Stone-Gross et al. analysed the Spamdot forum, studying the tightly connected community of buyers and sellers that 
were active on it~\cite{stone2011underground}. Onaolapo et al. showed that cybercriminals active on such forums actively look for free samples of stolen credentials, and assess their quality before making a purchase~\cite{onaolapo2016happens}.

Credit card information is generally divided into three groups: \textit{credit card numbers}, \textit{dumps}, and \textit{fullz}~\cite{holt2010exploring}. Credit card numbers (also known as \textit{``CVV''}) include at least the information printed on the card, that is, actual credit card number, cardholder name, expiration date and security code \textit{CCV2} on the back of the card (not to be confused with CVV), and sometimes, the billing address and phone number. 
Dumps denote information from the tracks on the magnetic stripe of a card. These data are required to clone physical credit cards. Fullz provide further information on the cardholder including, for example, date of birth or social security number~\cite{holt2010exploring}. 

On the prices sought for products on carding forums, Shulman~\cite{shulman2010underground} states that, in 2006, credit card numbers were traded for \$1-\$25 each. Only two years later, credit card numbers were available for \$0.06. Shulman~\cite{shulman2010underground} mentions three reasons that account for this decline: CVVs are becoming a commodity, monetizing information is more difficult and credit cards are beset by stolen online credentials. In the April 2015 report on Internet security, Symantec~\cite{symantec2015} indicates a price range of \$0.50-\$20 for CVVs. These rates, on the one hand, confirm that there are details sold at low prices and, on the other hand, show that there are still cards sold for \$20.

Sood and Enbody~\cite{sood2013crimeware} provide a more detailed estimation of rates charged per credit card number. Numbers from the USA cost \$4-\$10 on average, from Canada \$5-\$7 and from the UK \$6-\$8. Classified according to credit card types, a classic or standard credit card number from the USA or Canada costs \$8-\$10, a gold card \$15-\$20 and an Amex \$6-\$10. Classic and Amex cards are the cheapest in the listing of Sood and Enbody~\cite{sood2013crimeware}. Nevertheless, they are still more expensive than the lower limits of their quoted price range (\$4/\$5).  
However, it is not ideal that these rates have not been observed but estimated and it remains unclear 
on which basis they have been calculated.

Reasons for price differences include the types of cards and countries of origin, as already mentioned, in addition to the rarity and the quantity of the products to be purchased~\cite{hutchings2014crime}. Discounts on purchases of large card quantities lower the price per item. Furthermore, cards with more personal information available, with high balances and extended expiration dates and freshly acquired cards tend to be more expensive~\cite{ablon2014markets}.

Since there is no sound information available on the product proportions on carding forums so far, this paper provides some insights into that. Existing literature does not state reliable prices for dumps or fullz. As cards containing more information tend to be more expensive than those with less information, however, we infer and hypothesise (\hyp{1}) that prices for fullz are higher than those for credit card numbers. For dumps, no hypothesis can be formulated derived from existing research literature. Due to the effort needed to monetise the information, low prices would be expected. However, once copied and successfully used to conduct a transaction, such a clone might be a lucrative means of payment. Also, the efforts necessary to steal the data (e.g. skimming dumps in a restaurant or collecting data on fake Internet sites) does not give a clear indication of the expected price differences between dumps and CVVs.

\subsection{Seller prolificacy}

Generally, there are several types of participants on the forums: sellers, buyers, intermediaries, mules, administrators, and others. These roles are not mutually exclusive; sellers may simultaneously be buyers. Although the total number of participants is in unclear, Ablon et al.~\cite{ablon2014markets} argue that, based on expert interviews and literature review, the total number of participants on the forums is likely to rise. The increasing spread of different marketplaces and forums would facilitate access to one of them. At least from a historic perspective, Christin~\cite{christin2013traveling} confirmed this growth of participants on underground platforms as he observed a linear increase of sellers during his half-year analysis of Silk Road, a large underground marketplace. In the aftermath of Silk Road's take-down in 2013, the number of sellers on competitor{-} and newcomer-marketplaces has substantially increased, surpassing the original number of sellers on Silk Road~\cite{soska2015measuring}.

In terms of geographic location, forum users come from all over the world. Regarding the sellers, De Carbonnel~\cite{decarbonnel2013} claims that Russian participants deliver the best quality, while participants from China, Latin America and Eastern Europe are the leaders with respect to quantity. These geographic patterns, however, vary depending on the types of forums and the services provided. An analysis of a marketplace offering SEO services locates the sellers mainly in India, Bangladesh, and the USA~\cite{farooqi2015characterizing}.

Examinations in relation to sales quantity reveal substantial differences in seller prolificacy. Farooqi et al.~\cite{farooqi2015characterizing} identified an ``insider ring'' composed of several top sellers. This means that a small number of traders account for a large proportion of traffic on the marketplace. One common characteristic is that they joined the community very early and are frequent visitors to the pages. Christin~\cite{christin2013traveling} agrees on the existence of several long-time sellers but also reports on a continual ``come and go'' of sellers. It is unclear, however, whether they leave the community after having made sales or due to unsuccessful attempts.

In terms of seller prolificacy, Motoyama et al.~\cite{motoyama2011analysis} analysed the records of 6 closed forums and concluded that 10\% of the sellers are responsible for 40\%-50\% of the goods traded. More generally expressed, D{\'e}cary-H{\'e}tu and Lepp{\"a}nen~\cite{decary2013criminals} reason that some sellers are more effective than others. Their conclusion is based on counting of advertisements of sellers on one underground forum. However, it is doubtful that counting ads is the right approach of quantifying success. Moreover, the analysis of several forums instead of one might have produced more reliable results.

It is common ground among crime scientists that crime is distributed neither randomly nor evenly~\cite{felson2010crime}. That implies a small group accounts for more offenses than its expected share would be. As earlier stated, studies on marketplaces suggest the presence of some highly prolific users. We hypothesise (\hyp{2}) that on active carding forums, a small number of traders are responsible for a large proportion of traffic.

\subsection{Seller specialisation}

Looking at the products sold per seller, several studies found evidence of specialisation amongst sellers. Derived from literature review and expert interviews, Kraemer-Mbula et al.~\cite{kraemer2013cybercrime}, for example, promote an ecosystem perspective to understand the actions of underground traders. Comparable to the legitimate business community, underground ecosystems includes actors that compete against each other, targeting competitive advantage. They try to reach this advantage by specialising in a particular type of product~\cite{kraemer2013cybercrime}.

By applying a framework of social organisation, Holt~\cite{holt2013exploring} identified specialisation on underground forums too. While one third of sellers offered various products, two thirds focused on only one product category. As the Symantec report~\cite{symantec2015} illustrates, there are perpetrators specialising in writing viruses, in distributing malware or in monetising credit cards, for example. In recent years, Symantec has observed an increasing professionalisation in all aspects in the underground economy. Their findings are supported by research literature. Sood and Enbody~\cite{sood2013crimeware} also identified specialisation as a trend in underground markets. They argue that these markets are increasingly accessible to people with various technical skills. Hence, there is a division of labour due to differing skills. While analysing seller characteristics on black marketplaces, Soska and Christin~\cite{soska2015measuring} discovered numerous specialised sellers, though there was a notable number of vendors selling different products as well.

These findings indicate that specialisation is present in the underground ecosystem as in the legitimate business world. Hence, we hypothesise (\hyp{3}) that specialisation is also discernible on carding forums, that is, most of the traders sell only one product type.
 
What does that mean in terms of product prices? We did not find any association between specialisation and product prices in existing literature. Resorting to economic theories~\cite{smith1937inquiry}, there are long-established economic ``laws'' that basically state that concentrating on one production task leads to a higher efficiency at that particular task. This efficiency enables an increase in production compared to unspecialised suppliers. Due to such economies of scale, products and services can be offered at significantly lower costs and prices can be cut. Applying this to traders on carding forums, we expect cost{-} and price-reducing effects when sellers specialise in trading of a single product category (due to the economies of scale). Thus, we hypothesise (\hyp{4}) that specialised traders sell their products at lower prices than unspecialised traders.

\subsection{Seller reputation}

One key aspect in the underground economy is reputation~\cite{ablon2014markets, decary2013criminals, motoyama2011analysis}. A reputable seller is more likely to be trusted 
and thus more likely to engage in trades and to complete transactions. On forums, reputation is usually established by positive customer feedback. Buyers may rate their sellers by giving positive ratings if the ordered products have been successfully delivered, and negative ratings if the seller has not delivered and was rather a ripper. Consequently, a seller's positive reputation score presents his/her threads in a more credible light, and these sellers have a higher chance of acquiring multiple customers~\cite{holt2013exploring}.

However, the effort to establish baseline reputation appears to be laborious. Before half of traders receive their first positive feedback, for example, they write approximately sixty posts~\cite{motoyama2011analysis}. In this case, the reputation process is intrinsically peer-driven. Sellers are dependent on recommendations by buyers. Sometimes, forum administrators provide a vetting process, often in addition to the peer-driven process and often with intransparent criteria. In those cases, entry costs are relatively high and access to higher tiers is tight~\cite{ablon2014markets}.

The emphasis on reputation and trust means that it is indispensable for competitive forums to have a well-functioning reputation system. Again, since the above-mentioned findings are widely based on expert interviews and therefore remain relatively vague, the actual status regarding currently running carding forums is not known. Since trades on carding forums depend on relationships between mutually distrustful parties, we argue that trust is even more important than in legitimate trades. In the event of an unsuccessful deal, the parties hardly have any legal remedies and countermeasures available, except for a negative reputation rating. We thus hypothesise (\hyp{5}) that the carding forums to be analysed have working reputation systems that are at least as sophisticated as those of legal marketplaces, for instance eBay. This expectation applies only to open forums where everybody can participate.

As discussed, the efforts needed for gaining trust are extensive. A consequence might be that sellers concentrate on establishing reputation on one specific forum instead of several forums. It is therefore not expected that sellers are present on multiple forums. By assuming this, we support Motoyama's et al.~\cite{motoyama2011analysis} expectation of non-existing multiple accounts. In contrast, we disagree with Ablon et al.~\cite{ablon2014markets} who argue without providing any reasons that sellers would advertise on multiple marketplaces. We hypothesise (\hyp{6}), therefore, that the vast majority of actors are not operating on more than one forum.

\section{Methodology}
\label{sec:methodology}

In this section, we describe our data collection approach. We collected names of underground forums from various sources, and selected 5 forums that matched our selection criteria for examination. Data spanning a period of three months was collected from the forums, and we tested our hypotheses on the data.     

\subsection{Forum search}

The first step of the examination is the forum search. We took the following steps to find carding forums: First, we collected names of forums that were mentioned by research literature. Second, we carried out searches via Google. Third, we used other search engines and information pages, namely \texttt{Onion.city} search via Tor network, \texttt{webstatsdomain.org} and ``The Hidden Wiki.'' Finally, we searched forums that we already found for references to other forums. In the latter case, we adopted the method of snowball sampling~\cite{biernacki1981snowball}. The only selection criterion at this point was that, due to the authors' language abilities, the forums had to be at least partly in English or in German.

By this means, we found 25 forums, 15 of them via Google. The 25 forums are listed in Table~\ref{tab:discoveredforums}. The forum names mentioned in existing literature research were of little use since all the mentioned forums had already been shut down. We found five forums through other forums, and five from listings and other search pages. Although numerous forums were listed, most of them did not exist anymore. 
Two of the forums discovered during the first search in February 2015 were shut down at the beginning of the analysis in June 2015. Notwithstanding, the carding underworld seems to be dynamic. We found one active forum containing posts dating back to 2008.

\begin{table}[thp]
\centering
\begin{tabular}{|l|l|}
\hline
\textbf{Forum name} & \textbf{Forum address (http://...)} \\ \hline \hline
Agoraforum & \url{lacbzxobeprssrfx.onion/} \\ \hline
\textbf{Altenen} & \url{www.altenen.com} \\ \hline
\textbf{Crdpro} & \url{crdpro.su} \\ \hline
\textbf{Crimenetwork} & \url{crimenc5wxi63f4r.onion} \\ \hline
Cardingforum & \url{www.cardingforum.org} \\ \hline
Hackingforum & \url{hackingforum.ru} \\ \hline
Unixorder & \url{www.unixorder.com} \\ \hline
Crdclub & \url{crdclub.ws} \\ \hline
Carderscave & \url{www.carderscave.ru} \\ \hline
Infraud & \url{infraud.cc} \\ \hline
Lampeduza & \url{lampeduza.so} \\ \hline
Blackstuff & \url{www.blackstuff.net/forum.php} \\ \hline
Bus1Nezz & \url{bus1nezz.biz} \\ \hline
Cardingmafia & \url{www.cardingmafia.ws} \\ \hline
\textbf{Bpcsquad} & \url{www.bpcsquad.com} \\ \hline
Procarder & \url{www.procarder.ru} \\ \hline
Cardersforum & \url{www.cardersforum.se/} \\ \hline
Crimes & \url{crimes.ws/} \\ \hline
Carderbase & \url{carderbase.su} \\ \hline
Carder & \url{carder.me} \\ \hline
Darkstuff & \url{www.darkstuff.net} \\ \hline
Coinodeal & \url{coinodeal.com} \\ \hline
\textbf{Tuxedocrew} & \url{www.tuxedocrew.biz} \\ \hline
Privatemarket & \url{privatemarket.us} \\ \hline
Omerta & \url{omerta.cm} \\ \hline
\end{tabular}
\caption{Names and web addresses of discovered forums. The ones we focused on are 
highlighted in boldface text.}
\label{tab:discoveredforums}
\end{table}

Besides forums, we discovered more than two dozen stores (e.g.~\textit{Globalcards} and~\textit{Dexter}, offering mainly credit card numbers). We did not include these single-vendor marketplaces in our analysis since they differ significantly from forums. They do not gather multiple sellers, they have no reputation systems, and users normally do not communicate with each other. Hence, they do not meet our interest in interactions between forum members.

To narrow down the analyses, we chose five out of the 25 forums (see Table~\ref{tab:threadsandproductsperforum}) for detailed examination: \textit{Altenen}, \textit{Crdpro}, \textit{Crimenetwork}, \textit{Bpcsquad}, and \textit{Tuxedocrew}. The first three are the largest forums we found (as measured by number of posts) and should thus be the most fruitful ones. We excluded \textit{Agoraforum} despite possessing the greatest number of posts, because 99\% of its posts are requests for referral links for registration on Agora Marketplace. Tuxedocrew is included as it is one of the smallest forums and has existed for around two years. Thus, it is not entirely new and it might provide interesting insights when its content is compared to that of larger forums. Finally, we chose Bpcsquad since it is a medium-sized forum. It is remarkable to note that  it is the largest one of the very new forums. To sum up, our selection criteria are forum size, founding date, and, to a lesser degree, content. These criteria should ensure a good mix. Altenen, Bpcsquad and Tuxedocrew are mainly or exclusively in English, Crimenetwork in German and Crdpro half in English, half in Russian.

\subsection{Temporal sampling}

In order to have a comparable time-coverage of all five forums, we monitored activity on them over a period of three months, specifically from April to June 2015. This means that a snapshot was made by the end of June and data of the previous three months was collected. This three-month period was determined by the largest forum, Altenen. 

This limitation to three months meant that no full activity-record could be recorded. Furthermore, it introduced the risk of catching three ``special'' months instead of a whole year's coverage. However, we argue that the current situation is of interest and not the past, and that three months are still more advantageous than shorter periods. Moreover, the collected data showed that a substantial volume of posts can be captured in three months, especially from the larger forums. Admittedly, a longer period would be beneficial for the smaller forums. In summary, a consistent and thus comparable time period is favoured over a larger number of posts from small forums.

\subsection{Data collection}

We collected information on the numbers of members and posts, content, forum-accessibility, languages, and founding dates. For the selected forums, threads where potential sellers advertise their products were collected. A systematic review of the entirety of these forums was not possible. No activity records, copies of databases or web crawler services were available. The analysis was effected from the user's perspective. For instance, we did not analyse private messages used to arrange and complete trades. Nevertheless, this method provides an enlightening snapshot of the current carding situation. Where necessary and possible, we set up login credentials to gain wider access to the forums.

Threads published between April and June 2015 were collected for further analysis. Ads that were created before the three-month observation period were not collected. However, it is likely that older threads were still successfully promoting products and generating sales. Therefore, we captured older threads in cases where an activity in the form of answer postings or vouchings during the three months was registered. Such activity suggests that deals had taken place. Indeed, it was crucial to consider such older threads since it was expected that long-established insider rings existed on the forums, as pointed out by Farooqi et al.~\cite{farooqi2015characterizing}.

The threads usually describe the advertised products and their prices. Whenever an unspecified price range was indicated in an ad, we chose the lowest price for analysis. Calculating the mean value may distort the picture presuming that, for example, if only one high-priced gold card is offered in addition to many low-priced standard cards. In cases where various products were advertised in a single thread, each entry was considered equally.

To keep the focus on carding, we limited the spectrum of investigation to typical financial cybercrime related data: credit card numbers (CVVs), dumps, fullz, PayPal-credentials and Western Union (WU) payment transfers. We excluded other carding-related services such as ordinary online store credentials or monetisation-services.

In order to operationalise ``traffic'' on the forums, as necessary for hypothesis~\hyp{2}, D{\'e}cary-H{\'e}tu and Lepp{\"a}nen ~\cite{decary2013criminals} counted advertisements as indicators. In our view, however, the consideration of vouchings would be more promising to obtain an accurate impression of a seller's ``performance.'' Vouchings are evident signs that successful transactions have been made. Yet, since probably not every buyer vouches for the seller, counting the number of vouchings tends to underestimate the traffic. Conversely, there might be rippers vouching for each other without having made any transaction. Since Farooqi et al.~\cite{farooqi2015characterizing} and Christin~\cite{christin2013traveling} also relied on vouchings and member feedbacks, using them to calculate revenues, counting vouchings seems to be an appropriate method.

To determine whether users are specialised in one product category, as a prerequisite to be able to test~\hyp{3} and~\hyp{4}, we checked their personal profile sites. These pages display complete lists of all threads and posts written by the corresponding users. This method enabled us to see whether multiple products were advertised. The definition of specialisation is relatively strict. For instance, if users sold credit card numbers and PayPal-credentials, we did not consider them as being specialised. Only very narrowly related categories, for instance credit card numbers and fullz, were treated as identical product categories in this respect.

Hypothesis~\hyp{6} requires us to determine whether the same users are present on several forums. We carried out searches for users throughout the selected forums, and compared their identity details. Where applicable, these are username, email address, ICQ-number and Yahoo-ID. These details were collected from the postings and the users' profile pages.

Finally, we gathered information on the reputation system of each forum from various sources. Depending on the forum, these are the FAQs, specially installed forum threads, terms and conditions or customer information sites. Also own observation and interpretation were employed to grasp how the reputation systems functioned.

In total, we collected 388 threads. They advertised 987 individual products in total, that is, on average, each thread promoted 2.5 individual products (e.g. CVV USA Classic). The figures for the individual forums are in Table~\ref{tab:threadsandproductsperforum}.

\begin{table}[thp]
\centering
\begin{tabular}{|l|r|r|}
\hline
\textbf{Forum} & \textbf{Threads} & \textbf{Individual products}  \\
\hline
Altenen & 206 & 431 \\
\hline
Crdpro & 57 & 270 \\
\hline
Crimenetwork & 96 & 136 \\
\hline
Bpcsquad & 25 & 130 \\
\hline
Tuxedocrew & 4 & 20 \\
\hline
\end{tabular}
\caption{Threads and individual products per forum}
\label{tab:threadsandproductsperforum}
\end{table}

\subsection{Analytical strategy}

Monitoring of the forums required a combined methodical approach. We analysed the content we collected both qualitatively and quantitatively. Qualitative analysis was applied for content categorisation and analysis of reputation, while quantitative analysis was applied for comparisons of products and prices and determination of traffic per seller. 

To prepare the data for analysis, a categorisation of thread content was necessary. This procedure required a qualitative research approach and was thus done manually in Excel using content analysis method. We stuck to clear coding rules in order to avoid subjective and inconsistent categorisations. Some approximate categories were already provided by existing research literature (e.g. ``credit cards''). However, these categories were somewhat too coarse and further sub-categories had to be created (e.g. ``CVV'' or ``dumps''). Hence, 
The categorisation process is a combined product of deductive practice (assigning content to given categories) and inductive practice (building new categories based on content)~\cite{kluge1999empirisch}. Yet, it is important to ensure that the categories do not become too small and thus render subsequent quantitative calculations impossible. For example, dumps are sometimes advertised divided into \textit{track 1} or \textit{track 2} dumps. Since the aim of the categorisation is to obtain meaningful product categories, we avoided such fine distinctions. Therefore, credit card data was coded according to product category, country and product type (e.g. CVV, UK, gold). Visa and Mastercard details were not explicitly differentiated since they are usually treated interchangeably by the traders.

Next, we imported the data into SPSS software for statistical analyses. At first, we ran general frequency calculations. In order to test~\hypl{2}, the number of traders in relation to the generated traffic was quantitatively computed in the shape of a Lorenz curve. For~\hyp{3}, the frequencies of the specialised users were compared to the unspecialised ones. \hyp{1}~and \hyp{4} required the application of inference statistics. Since their price distributions resembled Poisson rather than a normal distribution, we performed Mann-Whitney-U tests  to test whether there were significant differences between the values. Regarding \hyp{5}, the reputation systems of the forums were evaluated qualitatively. To assess the degree of sophistication in relation to legitimate marketplaces, we compared them to eBay's system. Strictly speaking, eBay is not a forum and thus not fully comparable. However, we found no suitable legitimate large-scale forum set up to enable trading. Furthermore, eBay comes close to the system of advertising, trading, and buying exercised on forums. 
Finally, in order to examine whether sellers operated on more than one forum (\hyp{6}), we reproduced and interpreted the proportions of multiple representations across all forums.

We carried out all these calculations both for the entirety of the threads and for each forum separately. Therefore, the results (Section \ref{sec:analysis}) are reported in aggregated form, and where applicable, and if enough cases are available, for every individual forum.

\section{Data Analysis} 
\label{sec:analysis}

In this section, we describe our analyses of the five selected forums, and our findings.

\subsection{Overview}

We discovered 25 forums, out of which we selected five forums for analyses. In this section, we describe general attributes of the 25 discovered forums. The attributes are name, members, total posts, accessibility, languages, and founding date.

\desc{Name}~The names and full website addresses of the discovered forums are listed in Table~\ref{tab:discoveredforums}. They have a wide array of top-level domains, for example~\url{.com},~\url{.ws} (Samoa), and~\url{.so} (Somalia). The locations that the forums really operate from are usually unclear, and apart from two German-speaking platforms, cannot be derived from the forum content.

\desc{Members}~As of June 2015, the smallest forum had 1,100 members, while the largest had 148,800 members. A comparison between the first search in February 2015 and the second in June 2015 revealed some substantial increases in members. Cardersforum, for example, grew from 44,200 to 45,700 members (3.4\% increment), and Cardingmafia grew from 98,700 to 121,500 members (23.1\% increment). Altenen, already a large forum in February, was more than twice as large four months later (from 60,700 to 148,800 members, 145.1\% increment).

It is not clear how many of the members on each forum were actually contributing. Two forums indicated in their forum statistics that only a fraction of members were really active. On Altenen, these were 38,300 of its 148,800 members (25.7\%); on Carder, these were 1,500 of 10,200 (14.7\%). Neither forum disclosed what ``active members'' actually meant.

\desc{Total posts}~The number of posts varied between 150 and 15,778,599. Besides advertisements, the posts comprised mainly answers to advertisements or contributions to discussions. Typical answers to offers for sale are ``interested please contact me'' or ``made deal and worked.'' In line with the increase in number of members, the number of posts also increased between the two searches (e.g. Cardingmafia from 31,900 to 37,600 posts, 17.9\% increment). On Altenen, the number of posts doubled (from 607,100 to 1,265,500 posts, 108.5\% increment).

\desc{Accessibility}~Nine forums were completely open. This means that everybody could access them for free or even without registration. More than half of the forums had private VIP areas that required special registration to join. Access to these areas usually required a recommendation or an invitation by other members. Three forums charged registration fees, \$50 in the case of Lampeduza, and \$100 in the cases of Infraud and Omerta respectively.

\desc{Languages}~The forums were mostly in English and Russian, with two in German. Some forums contained international sections in various languages. However, the number of posts in such sections were consistently small.

\desc{Founding date}~We estimated the founding dates of the forums from the oldest posts found on them, mostly in the introduction or announcement sections. These are not necessarily the founding dates as older posts might have been deleted in the meantime. In addition, forums might have been shut down and reopened under another name (e.g. Crdpro was formerly Carderpro). We estimated founding dates between 2008 and 2014. A large number of the forums were launched in 2013.

In terms of size, the median number of members was 28,850, while the median number of posts was 58,150. Excluding the special case of Agoraforum, the number of posts per member varied between 0.1 (Privatemarket) and 18.6 (Crimenetwork). There are thus forums where only a fraction of the members post messages and there are some where members post numerous messages on average. However, these results have to be treated with caution as posts or members might have been deleted during the existence of the forums.

\subsection{Detailed analysis}

In this section, we present the results of hypotheses testing for the five selected forums.

\desc{Products and prices}~Table \ref{tab:productsandprices} presents the numbers, proportions (in \%) and prices (in US\$) per product category. CVVs are further divided by product type. Credit card numbers cost on average \$10, dumps and fullz more than \$30. PayPal credentials are advertised for \$3. Western Union payments of \$100 are sold for \$15. As hypothesised (\hyp{1}), prices for fullz are higher than those for credit card numbers (CVV: mean = 10.08, median = 10.00; fullz: mean = 31.82, median = 30.00). The difference is statistically significant (Mann-Whitney-U = 4011, z = -12.86, p~\textless~0.01).

\begin{table}[thp]
\centering
\begin{tabular}{|l|r|r|r|}
\hline
\textbf{Products} & \textbf{Number} & \textbf{Proportion (\%)} & \textbf{Price (\$)} \\ \hline
\textbf{CVVs} & 465 & 47.1 & 10.08 \\ \hline
\textit{Classic} & \textit{98} & \textit{9.9} & \textit{9.93} \\ \hline
\textit{Gold} & \textit{14} & \textit{1.4} & \textit{16.86} \\ \hline
\textit{Amex} & \textit{66} & \textit{6.7} & \textit{12.34} \\ \hline
\textit{others} & \textit{16} & \textit{1.6} & \textit{13.00} \\ \hline
\textit{unspecified} & \textit{271} & \textit{27.5} & \textit{9.06} \\ \hline
\textbf{Dumps} & 234 & 23.7 & 34.52 \\ \hline
\textbf{Fullz} & 140 & 14.2 & 31.82 \\ \hline
\textbf{PayPal} & 133 & 13.5 & 3.01 \\ \hline
\textbf{WU (\$100)} & 15 & 1.5 & 15.00 \\ \hline
\textbf{Total} & 987 & 100.0 &  \\ \hline
\end{tabular}
\caption{Products and prices (mean) in total.}
\label{tab:productsandprices}
\end{table}

Examined per individual forum, prices of CVVs do not vary substantially, those for the other products show considerable variation. Besides price differences, the product proportions alsi vary per forum. On Altenen, for example, dumps have a share of 8\% of the products analysed. On Crdpro, their proportion is 52\%. Yet the absolute numbers are partly very low and the values may thus lack reliability. 

The prices per product category depend widely on the effective composition of these categories, that is, on the relative frequencies per product type and country of origin. Since the absolute numbers are too low to display these values for each forum, Table~\ref{tab:productsandpricespertype} shows them summarised across all forums. Only product types consisting of at least 10 cases are considered. There is indeed substantial variation between different product types and countries. Amex and Gold cards are consistently more expensive than classic VISA or Mastercard cards. US products are the cheapest, while European products tend to be more expensive.

\begin{table}[tb]
\centering
\begin{tabular}{lrlrr}
\hline
\multicolumn{3}{l}{} & \textbf{Number} & \textbf{Price (\$)} \\ \hline \hline
\textbf{CVV} & \textbf{Australia} & \textbf{Classic} & 12 & 12.75 \\ \hline
\textbf{} & \textbf{Canada} & \textbf{Amex} & 10 & 14.20 \\ \hline
\multicolumn{2}{l}{\textbf{}} & \textbf{Classic} & 14 & 10.92 \\ \hline
\textbf{} & \textbf{UK} & \textbf{Amex} & 10 & 15.10 \\ \hline
\multicolumn{2}{l}{\textbf{}} & \textbf{Classic} & 17 & 11.94 \\ \hline
\textbf{} & \textbf{USA} & \textbf{Amex} & 25 & 7.02 \\ \hline
\multicolumn{2}{l}{\textbf{}} & \textbf{Classic} & 31 & 5.46 \\ \hline
\textbf{Dumps} & \textbf{Canada} & \textbf{Classic} & 14 & 31.43 \\ \hline
\multicolumn{2}{l}{\textbf{}} & \textbf{Gold} & 12 & 45.25 \\ \hline
\textbf{} & \textbf{EU} & \textbf{Classic} & 20 & 41.75 \\ \hline
\multicolumn{2}{l}{\textbf{}} & \textbf{Gold} & 20 & 58.50 \\ \hline
\textbf{} & \textbf{USA} & \textbf{Classic} & 29 & 19.17 \\ \hline
\multicolumn{2}{l}{\textbf{}} & \textbf{Gold} & 27 & 30.93 \\ \hline
\end{tabular}
\caption{Products and prices (mean) per product type with at least 10 cases.}
\label{tab:productsandpricespertype}
\end{table}

\begin{figure*}[thp]
  \begin{center}
    \subfloat[All forums.\label{fig:trafficpertradersall}]{\includegraphics[width=0.5\textwidth]{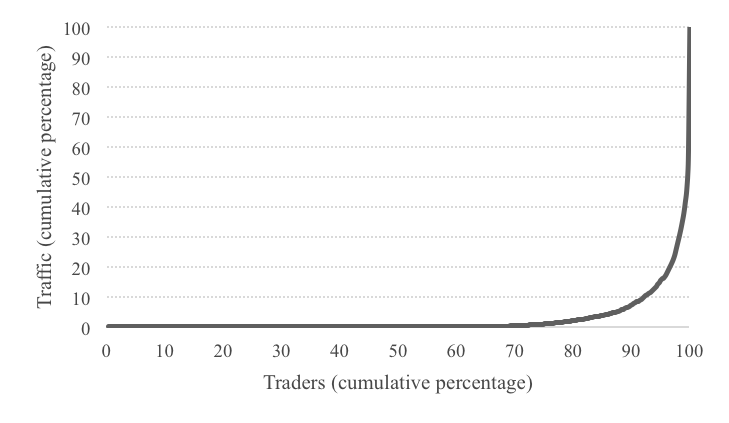}}
  \subfloat[Altenen forum.\label{fig:trafficpertradersaltenen}]{\includegraphics[width=0.5\textwidth]{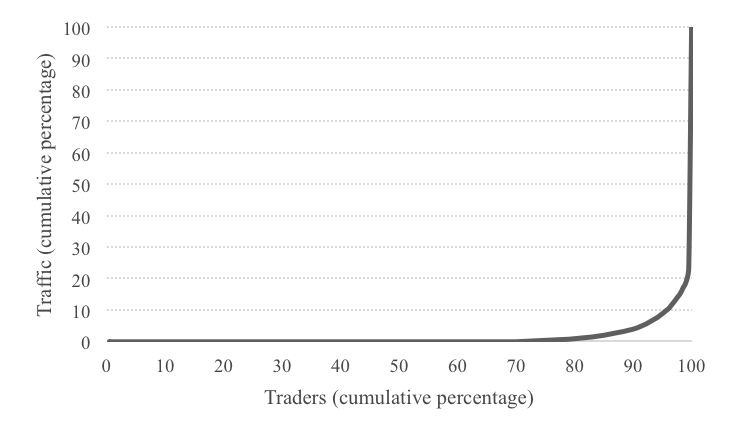}}
\end{center}
  \caption{Lorenz curves indicating cumulative percentage of traders against cumulative percentage of traffic on all forums, and Altenen forum.}
\end{figure*}

\desc{Seller prolificacy}~The products are not sold evenly throughout all sellers on the forums. The Lorenz curve in Figure~\ref{fig:trafficpertradersall} shows that around 70\% do not generate any obvious traffic as seller, whereas a single user generates 44\% of all traffic. This user joined Altenen in summer 2014 and sells CVVs of various countries.

\begin{figure*}[t]
  \begin{center}
  \subfloat[Crdpro forum.\label{fig:trafficpertraderscrdpro}]{\includegraphics[width=0.5\textwidth]{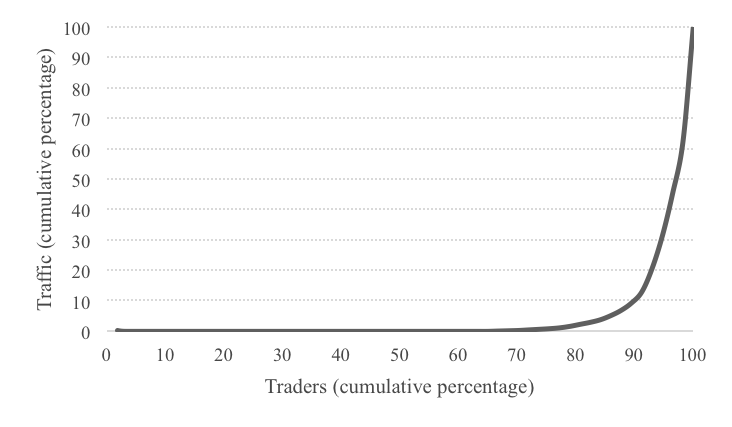}}
  \subfloat[Crimenetwork forum.\label{fig:trafficpertraderscrimenet}]{\includegraphics[width=0.5\textwidth]{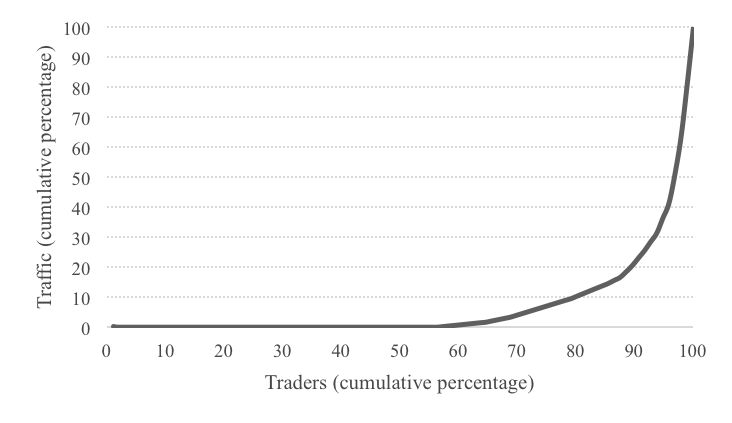}}
\end{center}
  \caption{Lorenz curves indicating cumulative percentage of traders against cumulative percentage of traffic on Crdpro and Crimenetwork forums.}
\end{figure*}

Disentangling the individual forums from the total results in similar pictures (Figures \ref{fig:trafficpertradersaltenen}, \ref{fig:trafficpertraderscrdpro}~and \ref{fig:trafficpertraderscrimenet}). Altenen has the most unequal distribution. Crimenetwork's distribution is not as extreme but still far from being equal. Bpcsquad and Tuxedocrew have too few vouchings to calculate a Lorenz curve. However, Tuxedocrew has posts dating back to 2013 that still receive vouchings but only in small numbers. Crdpro does not diverge from these distributions.

Considering the total amount of traffic, however, it is striking that the figures are low, both in the English and in the Russian speaking part. Although Crdpro has 17 times as many users as Bpcsquad, for example, it produced only twice as many advertising threads during the time of observation. Detailed analysis revealed that a substantial number of these threads contain links to shops. Furthermore, there have been no new entries in the two VIP areas since 2013 and the forum appeared to have been disconnected during some summer months in 2013.

In terms of products, the top three sellers on Altenen sell CVVs and WU payments, on Crdpro they sell dumps and on Crimenetwork again CVVs. As far as reputation is concerned, the high-profile sellers have usually high reputation ratings. To conclude this section, the hypothesis (\hyp{2}) that a small number of traders are responsible for a large proportion of traffic is accepted. However, we could not confirm the presence of an insider ring, as proposed by Farooqi et al.~\cite{farooqi2015characterizing}. Overall, only five out of the twenty most prolific users registered in the founding year of the according forum. The others joined later. However, it is possible that some sellers have more than one account and have thus several ``joining-dates.'' This possibility does not seem to be very likely, mainly due to the expected effort needed for establishing reputation for each account.

\desc{Seller specialisation}~In total, the majority of the users on the forums are not specialised (see Table~\ref{tab:specandunspecusers}), that is, most users sell more than one type of product and~\hypl{3} has to be rejected.

\begin{table*}[thp]
\centering
\begin{tabular}{|l|r|r|r|r|}
\hline
\multirow{2}{*}{} & \multicolumn{2}{c|}{\textbf{Specialised users}} & \multicolumn{2}{c|}{\textbf{Unspecialised users}} \\ \cline{2-5} 
 & \textbf{Number} & \textbf{Proportion (\%)} & \textbf{Number} & \textbf{Proportion (\%)} \\ \hline
\textbf{Altenen} & 64 & 31.1 & 142 & 68.9 \\ \hline
\textbf{Crdpro} & 41 & 73.2 & 15 & 26.8 \\ \hline
\textbf{Crimenetwork} & 24 & 25.0 & 72 & 75.0 \\ \hline
\textbf{Bpcsquad} & 15 & 60.0 & 10 & 40.0 \\ \hline
\textbf{Tuxedocrew} & 3 & 100.0 & 0 & 0.0 \\ \hline
\textbf{Total} & 147 & 38.1 & 239 & 61.9 \\ \hline
\end{tabular}
\caption{Number and proportion of specialised and unspecialised users per forum and in total.}
\label{tab:specandunspecusers}
\end{table*}

Regarding Crdpro, Bpcsquad and Tuxedocrew, the hypothesis would be true. However, a closer look on Crdpro reveals that its users generally sell a large variety of the same product category, instead, for instance credit cards from many different countries. This pattern is exactly the opposite of Crimenetwork's. Users on Crimenetwork usually sell different product categories but not various types within the same product category, for example only CVVs from Germany.

The results for~\hyp{3} raise the question about differences between specialised and unspecialised sellers. Building upon~\hyp{2} and taking into account the number of vouchings these users receive, no major differences are discernible. Regarding Altenen, for example, seven out of the twenty users with the most vouchings are specialised. This equals approximately the calculated specialisation rate of 31\%.

In terms of prices per product, there are some differences. CVVs, dumps, and PayPal-credentials advertised by specialised users are cheaper than those of unspecialised users; fullz are more expensive (see Table~\ref{tab:productcategories}). However, only the price difference for dumps is statistically significant at the 95\% confidence level (Mann-Whitney-U = 1643, z = -4.83, p~\textless~0.01). 
\hypl{4}~thus has to be partially rejected. 

That said, it is delicate to summarise product types because every type and country has its own price. A careful comparison would only contain a single product type. Hence, we made such a price comparison for US Classic CVVs, the most prevalent product type. The result shows no significant difference (Mann-Whitney-U = 54, z = -0.63, n.s.).

A striking aspect that Table~\ref{tab:productcategories} reveals is the distribution of advertised product categories among specialised and unspecialised users. Unspecialised users hardly advertise any dumps (7.5\% of all products). In contrast, dumps are almost half of the products (45.4\%) that specialised users advertise.

\begin{table*}[thp]
\centering
\begin{tabular}{|l|r|r|r|r|r|r|}
\hline
\multirow{2}{*}{} & \multicolumn{3}{c|}{\textbf{Specialised users}} & \multicolumn{3}{c|}{\textbf{Unspecialised users}} \\ \cline{2-7} 
 & \textbf{Number} & \textbf{Proportion (\%)} & \textbf{Price (\$)} & \textbf{Number} & \textbf{Proportion (\%)} & \textbf{Price (\$)} \\ \hline
\textbf{CVV} & 166 & 39.1 & 9.28 & 299 & 54.7 & 10.46 \\ \hline
\textbf{Dumps} & 193 & 45.4 & 32.56 & 41 & 7.5 & 42.61 \\ \hline
\textbf{Fullz} & 45 & 10.6 & 35.86 & 95 & 17.4 & 30.20 \\ \hline
\textbf{Paypal} & 21 & 4.9 & 1.99 & 112 & 20.5 & 3.17 \\ \hline
\textbf{Total} & 425 & 100.0 &  & 547 & 100.0 &  \\ \hline
\end{tabular}
\caption{Number, proportion, and price of product categories per specialised and unspecialised users.}
\label{tab:productcategories}
\end{table*}

\desc{Seller reputation}~We hypothesised (\hyp{5}) that carding forums have working reputation systems at least as sophisticated as those of legal marketplaces. The reputation systems are as follows:

\textit{eBay:} Buyers on eBay can leave feedback for a seller after a purchase and transaction ends. These ratings determine the ``feedback score.'' Positive feedback gives one point, neutral feedback does not change the score, and negative feedback subtracts one point~\cite{ebayfeedback}. As an additional protection measure, eBay refunds the purchase price in the event of a non-delivery~\cite{ebaymoneyback}. Furthermore, a seller protection service identifies high-risk buyers in order to avoid non-payments~\cite{ebaysellerprotection}.

\textit{Altenen:} Altenen's basic reputation system works in the same way as eBay's. A user's ``reputation power'' consists of the number of positive minus the number of negative feedback points. The median score of the observed sellers was 1. In addition, users are allowed to rate threads. Sellers also have the opportunity of paying \$50 to the ``Altenen buyer protection reserve'' that is used as backup payment service. In case of non-delivery, buyers receive their money back out of this fund. Altenen also offers  an escrow service that protects buyers from non-delivery and sellers from non-payment. In using that service, a buyer pays the money plus a transaction fee of \$5-\$30 to Altenen. Once the requested products are delivered, the amount is released to the seller.

\textit{Crdpro:} In theory, Crdpro has a feedback system identical to Altenen's. However, the system was disabled during our observation. As a consequence, apart from some long-standing members, all other members had neutral feedback scores. An escrow service was not provided.

\textit{Crimenetwork:} Crimenetwork's reputation system was based on ``likes,'' similar to Facebook. Members may ``like'' other members. Like-scores between 0 and 857 were recorded with a median of 36 likes. Crimenetwork's escrow service is comparable to Altenen's. A fee of 4\% of the purchase price is charged for successful transactions.

\textit{Bpcsquad:} As seen with Altenen and Crdpro, members may rate other members on Bpcsquad by giving positive, neutral or negative feedback. The scores ranged from 0 to 80, with a median of 0. Furthermore, there is a thread rating possibility. Bpcsquad also activated an escrow service but did not provide information on transaction fees.

\textit{Tuxedocrew:} Tuxedocrew's reputation system differed from those seen so far. Users could only rate threads but could not give any feedback for other users. Instead, the forum administrator could award users with special titles. The criteria that had to be met to receive these titles were not published on the forum. Tuxedocrew also offered an escrow service and charged a 15\% fee.

Overall, only Altenen's system appeared to be similarly elaborated as eBay's. However, it had an amateurish touch, especially the \$50 buyer protection reserve which is not able to cover substantial amounts. The other forums had fewer features than eBay and even those were not always working. \hyp{5}~is thus rejected. 

\desc{Presence of sellers on multiple forums} Finally, we examined whether users were present on several forums. This was done for all kind of sellers including high{-} and low-profile traders. In total, only six sellers were found trading on more than one forum, namely two on Altenen and Crdpro, two on Altenen and Bpcsquad, and two on Bpcsquad and Crdpro. A detailed analysis of these sellers showed that they were not high-profile but rather low-profile unsuccessful traders trying their luck on several platforms. \hypl{6}~is thus confirmed; concentration on a single forum was expected.

\section{Discussion}

In this section, we summarise our findings on the carding forums that we studied. The apparent lack of specialisation of forum users is also described. Finally, we highlight limitations of the study.

\desc{Summary of our findings}~The prices sought for the products offered on the forums lie within the range given by the reviewed literature. Dumps and fullz are relatively expensive; they are more 
than three times as expensive as credit card numbers (CVVs). This may be due to the effort needed to gain or monetise the data, the amount of information available, the higher rewarding potential, and differing demand and supply. Brison~\cite{brison2015} argues, for example, that dumps generally promise a higher payoff than CVVs. In contrast, CVVs are well-represented on the forums and thus seem to be available in abundance, which might push prices downwards. However, contrary to Shulman's assumption~\cite{shulman2010underground}, the prices of CVVs are still solid. Taking into account the large proportion of CVVs on the investigated forums, trading credit card numbers is presumably still a lucrative business. PayPal-credentials are well-represented on some forums as well, but so far do not seem to replace credit cards as the most attractive trading goods. Western Union money transfer services play only a marginal role on most of the forums. 

The products are advertised by sellers with varied success. Even though some users complete hundreds of transactions, most users do not sell anything at all. This means that the trading sections of the forums are profitable distribution channels for high-profile actors. 
This domination by a few traders implies that the forums are not typical forums characterised by mutual exchanging and participating users. In the carding world, there is somewhat a clash of prolificacy and -– arguably –- professionalism observable.

Referring back to the methodology part, counting of vouchings instead of ads, the latter~\cite{decary2013criminals} was probably more suitable to determine criminal performance. Some prolific sellers had only one ad but received dozens of vouchings. Counting of ads would have overlooked that.

Specialisation is not a key characteristic of sellers, even not of high-profile traders. Specialisation was observed mostly on Crdpro. This might be due to the high proportion of dumps sold on this forum. Dumps constitute almost half of the products sold by specialised users on Crdpro. Dealing with dumps appears to demand a higher degree of specialisation than dealing with only electronically obtainable products. Unlike CVVs or credentials, the acquisition of dumps requires a connection to the physical world. Therefore, perpetrators cannot stay in the underground cyberworld only. As a result, it might be costlier for unspecialised users to acquire dumps, thus forcing them to sell dumps at higher prices, which would confirm Smith's economic theory~\cite{smith1937inquiry}.

Yet the majority of sellers are not specialised. It could be argued that if they are apt or have valuable data sources, they know and distribute other types of illicit products and services. On the contrary, unsuccessful sellers try their luck with another product if it does not work with the first. These users, though, might as well be rippers. Advertising a large array of products might be done to give the impression of a prosperous seller, or they just try various products in the hope that somebody would engage in a trade eventually.

Overall, it is possible that the scope of analysis regarding specialisation was too narrow. There may be a specialisation in the underground world in larger terms where carding itself is already a specialisation. Another reason might be that carding is not as complicated as other cybercrimes like DDoS-attacks or large-scale spam campaigns. Regarding DDoS-attacks, for example, taking advantage of security vulnerabilities and manipulating compromised machines to send huge amounts of data may require more time and skills than stealing and trading credit card data. Therefore, it makes more sense to be specialised in those domains.

At least on the investigated forums, and given the available details, users are not present on more than one forum. It might thus be true that the effort needed to reach a certain reputation level deters users from establishing themselves on multiple forums, as Motoyama et al.~\cite{motoyama2011analysis} proposed. This effort could also be the reason why most users do not have any ratings at all, as the analysis showed. Another reason for this, however, might be the presence of rippers. Regarding users with high reputation and many vouchings, it is highly unlikely to find any rippers among them. Among users without any reputation scores and vouchings, the proportion of rippers could be large. It is interesting to note that an expert interviewed by Ablon et al.~\cite{ablon2014markets} estimates that around 30\% of all sellers are rippers.

In general, and if not stated otherwise, all our findings apply to all five examined forums. However, there are some differences. Sales on Crimenetwork are not distributed as extremely unevenly as on other forums, neither are there numerous specialised users present. Various people sell various goods. Crimenetwork is thus more forum-like in terms of mutual exchange and participation than the other forums. The high number of posts per member confirms this perception. In addition, the forum gives the impression of being well-maintained. It has a myriad of banned users and the administrators comment rigorously if users do not stick to the rules (e.g. in case a post does not fit into a thread).

Crdpro is the obvious opposite. Its best times were probably in the past. It appears to be in decline. It does not seem to be monitored by the administrators, there is no escrow service, the reputation system does not work, and there is in general not much traffic. It might be a question of time until the entire forum will be closed.

On Tuxedocrew, the smallest forum, there are only four recorded threads receiving any vouches. However, the number of vouchings are low and it is thus questionable how fruitful the business really is. What might be possible is that some recurring customers buy a lot and do not always vouch. A reason for the small size of this forum might be the high charges for the escrow service or, even more likely, the lack of a user-based reputation system. Only the administrator is able to assess other users, based on intransparent criteria. This might be too little to build trust among the users and to boost trade.

Bpcsquad and Altenen do not particularly diverge from the general findings. Bpcsquad is relatively small and the low number of ratings may denote unsuccessful deals. It is thus uncertain how strongly this forum will grow in the future. In contrast, the enormous increase in members on Altenen is impressive. Apparently, it is attractive to be part of this large community. Forums with numerous users usually have diverse products, and a multitude of potential buyers, that is, high supply and demand. Both platforms have reasonable and working basic reputation systems. Altenen provides an additional, arguably pseudo-protection measure.

\desc{Limitations}~We encountered a number of limitations during the study, and they are mentioned in this section. Firstly, the three-month period does not allow long-term conclusions. After all, due to the technique of considering the vouchings of this time period, older and often very profitable ads were included in the analysis. Secondly, the examination was carried out from a user's perspective. That is, no private messages could be studied. In addition, VIP sections on the forums had to be ignored. Thus, the findings of this study do not give a complete picture of the forums. Nevertheless, we gathered substantial amounts of data that allowed some analyses and conclusions. The third limitation concerns the internal validity of the data. It cannot be excluded that other investigators, for instance, law enforcement agents engage in trades on the forums for research and investigative purposes. This might bias the data. However, we do not consider this possibility a substantial threat.

Another threat to internal validity is the recorded product prices. The prices advertised are not necessarily the prices that buyers eventually paid. No post was found where the possibility of price negotiations was mentioned. Nevertheless, it cannot be excluded that users pay other, probably lower prices than those advertised.

Finally, given that the examined forums trade different goods or attract certain types of users, the findings are an artefact of the forums in question and do not represent the entire carding underworld. The results are only valid for the five forums we analysed. The selection of the forums is thus a threat to external validity. This limitation concerns all hypotheses but especially~\hyp{6} where an explicit cross-forum comparison was executed. In principle, this limitation was overcome by selecting five different forums based on various selection criteria.

\desc{Future work}~
Sellers were focus of this study. Future research should also consider their counterparts, the buyers. The reviewed literature did not cover buyers and they were also neglected in this paper. It might be useful to examine whether there are high-profile buyers and observe what 
they buy, and whether they resell the products, and to whom, if they resell. Regarding research design and methodology, a long-term or a follow-up study might be able to identify trends or confirm the patterns found in this study, respectively. 
Researchers could also consider engaging in trades and getting in touch with the traders, subject to ethical considerations. This method would allow researchers to collect more information on traders and gain better understanding of their roles within the fraud chain. These findings would help to shed more light on how to counter financial cybercrime in the future.

\section{Conclusion}

This paper presented an overview of 25 existing online carding forums and an in-depth analysis of five of these forums, covering a three-month period of monitoring. What differentiates this study from others is, first, we investigated real data instead of drawing conclusions based solely on existing literature or expert opinion, second, we examined active forums instead of closed forums, and third, we applied a low-level focus on products, prices and sellers. Our findings suggest that the market of carding forums is dynamic. However, it is not clear how promising the future of carding forums is, especially with the emergence of single-vendor stores which could imply that high-profile sellers would leave existing carding forums to open their own single-vendor stores.

\bibliographystyle{IEEEtran}
\bibliography{biblio}

\begin{thebibliography}{10}
\providecommand{\url}[1]{#1}
\csname url@samestyle\endcsname
\providecommand{\newblock}{\relax}
\providecommand{\bibinfo}[2]{#2}
\providecommand{\BIBentrySTDinterwordspacing}{\spaceskip=0pt\relax}
\providecommand{\BIBentryALTinterwordstretchfactor}{4}
\providecommand{\BIBentryALTinterwordspacing}{\spaceskip=\fontdimen2\font plus
\BIBentryALTinterwordstretchfactor\fontdimen3\font minus
  \fontdimen4\font\relax}
\providecommand{\BIBforeignlanguage}[2]{{%
\expandafter\ifx\csname l@#1\endcsname\relax
\typeout{** WARNING: IEEEtran.bst: No hyphenation pattern has been}%
\typeout{** loaded for the language `#1'. Using the pattern for}%
\typeout{** the default language instead.}%
\else
\language=\csname l@#1\endcsname
\fi
#2}}
\providecommand{\BIBdecl}{\relax}
\BIBdecl

\bibitem{ffauk2015}
F.~F.~A. UK, ``{Fraud The Facts 2015},''
  \url{http://www.financialfraudaction.org.uk/Fraud-the-Facts-2015.asp}, 2015,
  [Online: Accessed 23-February-2016].

\bibitem{gold2013identity}
S.~Gold, ``Identity crisis?'' \emph{Engineering Technology}, vol.~8, no.~10,
  pp. 32--35, November 2013.

\bibitem{allodi2015then}
L.~Allodi, M.~Corradin, and F.~Massacci, ``Then and now: on the maturity of the
  cybercrime markets (the lesson that black-hat marketeers learned),'' 2015.

\bibitem{afroz2013honor}
S.~Afroz, V.~Garg, D.~McCoy, and R.~Greenstadt, ``Honor among thieves: A
  common's analysis of cybercrime economies,'' in \emph{eCrime Researchers
  Summit (eCRS), 2013}.\hskip 1em plus 0.5em minus 0.4em\relax IEEE, 2013, pp.
  1--11.

\bibitem{yip2013forums}
M.~Yip, N.~Shadbolt, and C.~Webber, ``Why forums?: an empirical analysis into
  the facilitating factors of carding forums,'' in \emph{Proceedings of the 5th
  Annual ACM Web Science Conference}.\hskip 1em plus 0.5em minus 0.4em\relax
  ACM, 2013, pp. 453--462.

\bibitem{ablon2014markets}
L.~Ablon, M.~C. Libicki, and A.~A. Golay, \emph{Markets for Cybercrime Tools
  and Stolen Data: Hackers' Bazaar}.\hskip 1em plus 0.5em minus 0.4em\relax
  Rand Corporation, 2014.

\bibitem{motoyama2011analysis}
M.~Motoyama, D.~McCoy, K.~Levchenko, S.~Savage, and G.~M. Voelker, ``An
  analysis of underground forums,'' in \emph{Proceedings of the 2011 ACM
  SIGCOMM conference on Internet measurement conference}.\hskip 1em plus 0.5em
  minus 0.4em\relax ACM, 2011, pp. 71--80.

\bibitem{stone2011underground}
B.~Stone-Gross, T.~Holz, G.~Stringhini, and G.~Vigna, ``{The underground
  economy of spam: A botmaster's perspective of coordinating large-scale spam
  campaigns},'' in \emph{{USENIX Workshop on Large-Scale Exploits and Emerging
  Threats (LEET)}}, 2011.

\bibitem{onaolapo2016happens}
J.~Onaolapo, E.~Mariconti, and G.~Stringhini, ``What happens after you are
  pwnd: Understanding the use of leaked webmail credentials in the wild,'' in
  \emph{ACM SIGCOMM Internet Measurement Conference (IMC)}, 2016.

\bibitem{holt2010exploring}
T.~J. Holt and E.~Lampke, ``Exploring stolen data markets online: products and
  market forces,'' \emph{Criminal Justice Studies}, vol.~23, no.~1, pp. 33--50,
  2010.

\bibitem{shulman2010underground}
A.~Shulman, ``The underground credentials market,'' \emph{Computer Fraud \&
  Security}, vol. 2010, no.~3, pp. 5--8, 2010.

\bibitem{symantec2015}
S.~Corporation, ``2015 internet security threat report, vol. 20,'' technical
  report, Symantec Corporation, Tech. Rep., 2015.

\bibitem{sood2013crimeware}
A.~K. Sood and R.~J. Enbody, ``Crimeware-as-a-service—a survey of
  commoditized crimeware in the underground market,'' \emph{International
  Journal of Critical Infrastructure Protection}, vol.~6, no.~1, pp. 28--38,
  2013.

\bibitem{hutchings2014crime}
A.~Hutchings and T.~J. Holt, ``A crime script analysis of the online stolen
  data market,'' \emph{British Journal of Criminology}, p. azu106, 2014.

\bibitem{christin2013traveling}
N.~Christin, ``Traveling the silk road: A measurement analysis of a large
  anonymous online marketplace,'' in \emph{Proceedings of the 22nd
  international conference on World Wide Web}.\hskip 1em plus 0.5em minus
  0.4em\relax International World Wide Web Conferences Steering Committee,
  2013, pp. 213--224.

\bibitem{soska2015measuring}
K.~Soska and N.~Christin, ``Measuring the longitudinal evolution of the online
  anonymous marketplace ecosystem,'' in \emph{24th USENIX Security Symposium
  (USENIX Security 15)}, 2015, pp. 33--48.

\bibitem{decarbonnel2013}
A.~de~Carbonnel, ``{Hackers for hire: Ex-Soviet tech geeks play outsized role
  in global cyber crime},''
  \url{http://www.nbcnews.com/technology/hackers-hire-ex-soviet-tech-geeks-play-outsized-role-global-6C10981346},
  2013, [Online: Accessed 23-February-2016].

\bibitem{farooqi2015characterizing}
S.~Farooqi, M.~Ikram, G.~Irfan, E.~De~Cristofaro, A.~Friedman, G.~Jourjon,
  M.~A. Kaafar, M.~Z. Shafiq, and F.~Zaffar, ``Characterizing seller-driven
  black-hat marketplaces,'' \emph{arXiv preprint arXiv:1505.01637}, 2015.

\bibitem{decary2013criminals}
D.~D{\'e}cary-H{\'e}tu and A.~Lepp{\"a}nen, ``Criminals and signals: An
  assessment of criminal performance in the carding underworld,''
  \emph{Security Journal}, 2013.

\bibitem{felson2010crime}
M.~Felson and R.~L. Boba, \emph{Crime and everyday life}.\hskip 1em plus 0.5em
  minus 0.4em\relax Sage, 2010.

\bibitem{kraemer2013cybercrime}
E.~Kraemer-Mbula, P.~Tang, and H.~Rush, ``The cybercrime ecosystem: Online
  innovation in the shadows?'' \emph{Technological Forecasting and Social
  Change}, vol.~80, no.~3, pp. 541--555, 2013.

\bibitem{holt2013exploring}
T.~J. Holt, ``Exploring the social organisation and structure of stolen data
  markets,'' \emph{Global Crime}, vol.~14, no. 2-3, pp. 155--174, 2013.

\bibitem{smith1937inquiry}
A.~Smith, \emph{An Inquiry into the Nature and Causes of the Wealth of
  Nations}.\hskip 1em plus 0.5em minus 0.4em\relax Random House, 1937.

\bibitem{biernacki1981snowball}
P.~Biernacki and D.~Waldorf, ``Snowball sampling: Problems and techniques of
  chain referral sampling,'' \emph{Sociological methods \& research}, vol.~10,
  no.~2, pp. 141--163, 1981.

\bibitem{kluge1999empirisch}
S.~Kluge, ``Empirisch begr{\"u}ndete typenbildung: Zur konstruktion von typen
  und typologien in der qualitativen forschung,'' \emph{Opladen: Leske+
  Budrich}, 1999.

\bibitem{ebayfeedback}
``{How Feedback works},''
  \url{http://pages.ebay.co.uk/help/feedback/howitworks.html}, [Online:
  Accessed 23-February-2016].

\bibitem{ebaymoneyback}
``{eBay Money Back Guarantee},''
  \url{http://pages.ebay.co.uk/ebay-money-back-guarantee/}, [Online: Accessed
  23-February-2016].

\bibitem{ebaysellerprotection}
``{eBay Seller Protection},''
  \url{http://portal.ebay.co.uk/seller-protection/}, [Online: Accessed
  23-February-2016].

\bibitem{brison2015}
B.~Uri, ``{`Fullz', `Dumps', and more: Here's what hackers are selling on the
  black market},''
  \url{http://venturebeat.com/2015/02/08/fullz-dumps-and-cvvs-heres-what-hackers-are-selling-on-the-black-market/},
  2015, [Online: Accessed 23-February-2016].

\end{thebibliography}

\end{document}